\documentclass[final,5p,times]{elsarticle}

\usepackage{amssymb}
\usepackage{graphicx}
\usepackage{bm}
\usepackage{color}
\journal{Physics Letters B}

\begin{document}

\begin{frontmatter}

\title{Determination of the nuclear incompressibility from the rapidity-dependent elliptic flow in heavy-ion collisions at beam energies 0.4\emph{A} - 1.0\emph{A} GeV}

\author[1]{Yongjia Wang\corref{corr1}}\ead{wangyongjia@zjhu.edu.cn}
\author[2]{Chenchen Guo}
\author[1,3]{Qingfeng Li\corref{corr1}}\ead{liqf@zjhu.edu.cn}
\author[4]{Arnaud Le F\`evre}
\author[4]{Yvonne Leifels}
\author[4]{Wolfgang Trautmann}
\cortext[corr1]{Corresponding author}
\address[1]{School of Science, Huzhou University, Huzhou 313000, China}
\address[2]{Sino-French Institute of Nuclear Engineering and Technology, Sun Yat-sen University, Zhuhai 519082, China}
\address[3]{Institute of Modern Physics, Chinese Academy of Sciences, Lanzhou 730000, China}
\address[4]{GSI Helmholtzzentrum f\"ur Schwerionenforschung GmbH, D-64291 Darmstadt, Germany}

\begin{abstract}
\textbf{Background:} The nuclear incompressibility ($K_0$) plays a crucial role in understanding diverse phenomena in nuclear structure and reactions, as well as in astrophysics. Heavy-ion-collision measurements in combination with transport model simulations serve as important tools for extracting the nuclear incompressibility. However, uncertainties in transport models (or model dependence) partly affect the reliability of the extracted result.
\textbf{Purpose:} In the present work, by using the recently measured data of rapidity-dependent flows, we constrain the incompressibility of nuclear matter and analyse the impact of model uncertainties on the obtained value.
\textbf{Method:} The method is based on the newly updated version of the ultrarelativistic quantum molecular dynamics (UrQMD) model in which the Skyrme potential energy-density functional is introduced. Three different Skyrme interactions which give different incompressibilities varying from  $K_0$=201 to 271 MeV are adopted. The incompressibility is deduced from the comparison of the UrQMD model simulations and the FOPI data for rapidity-dependent elliptic flow in Au+Au collisions at beam energies 0.4\emph{A} - 1.0\emph{A} GeV.
\textbf{Results:} The elliptic flow $v_2$ as a function of rapidity $y_0$ can be well described by a quadratic fit $v_2=v_{20} + v_{22}\cdot y_0^2 $. It is found that the quantity $v_{2n}$ defined by $v_{2n}=|v_{20}|+|v_{22}|$ is quite sensitive to the incompressibility $K_0$ and the in-medium nucleon-nucleon cross section, but not sensitive to the slope parameter $L$ of the nuclear symmetry energy.
\textbf{Conclusions:} With the FU3FP4 parametrization of the in-medium nucleon-nucleon cross section, an averaged $K_0 = 220 \pm 40$~MeV is extracted from the $v_{2n}$ of free protons and deuterons. However, remaining systematic uncertainties, partly related to the choice of in-medium nucleon-nucleon cross sections, are of the same magnitude ($\pm 40$~MeV). Overall, the rapidity dependent elliptic flow supports a soft symmetric-matter equation-of-state.
\end{abstract}


\end{frontmatter}

The nuclear incompressibility ($K_0$) is defined as the derivative of pressure \emph{P} with respect to the density $\rho$, $K_0=9\left(\frac{\partial{P}}{\partial\rho}\right)|_{\rho=\rho_{0}}$, or the curvature of the energy per nucleon $E/A$ in nuclear matter at the saturation density ($\rho_0$), $K_0=9\rho^2\left(\frac{\partial^2E/A}{\partial\rho^2}\right)|_{\rho=\rho_{0}}$. The saturation density $\rho_0 \sim 0.16$ fm$^{-3}$ and the saturation energy $E_0 \sim -16$ MeV have been widely accepted and appear in textbooks. The equation of state (EOS) of symmetric nuclear matter can be expanded as $\frac{E}{A}(\rho)=E_0+\frac{K_0}{18}(\frac{\rho-\rho_0}{\rho_0})^2+...$, therefore, a more accurate value of $K_0$ means a better understanding of the EOS in the vicinity of the saturation density. As the knowledge of the EOS is essential for studying nuclear structure and reactions, as well as astrophysics, many attempts have been made to infer $K_0$ by using experimental data on the properties of nuclei (such as the giant monopole/dipole resonance and the nuclear masses and radii) or heavy-ion collisions (HIC) since 1960s.

Constraints on $K_0$ through comparing experimental data on nuclear properties (such as, the giant monopole resonance (GMR) energies in some nuclei, nuclear masses and charge radii) and theoretical model (as, e.g., the Skyrme-Hartree-Fock or relativistic mean-field extended by quasiparticle random-phase approximation) calculations have been summarized recently in Ref.\cite{Stone:2014wza}. In Ref.\cite{Khan:2013mga}, the authors studied the GMR energies of $^{208}$Pb and $^{120}$Sn, based on the constrained Hartree-Fock-Bogoliubov (CHFB) approach, and pointed out that $K_0$ varies in the region of $190<K_0<270$ MeV. However, different models offer a wide range of results for $K_0$ (see, e.g., \cite{Stone:2014wza,Khan:2013mga,Giuliani:2013ppnp} and references therein). Heavy-ion collisions provide the unique way to compress nuclear matter to high densities in the laboratory. They can serve as a powerful tool for studying the EOS of dense nuclear matter. However, the compressed nuclear matter exists only for a very short time (typically from several to several tens of fm/c, $10^{-23}$-$10^{-22}$ s), therefore, $K_0$ cannot be measured directly but only inferred from the comparison of experimental measurements with transport model simulations.

Since 1960s, several heavy-ion collision facilities became available, such as the BEVALAC at Berkeley and the NSCL at Michigan State University in US, the SIS of GSI at Darmstadt in Germany, the GANIL cyclotron at Caen in France, and the CSR at LanZhou in China. Extensive heavy-ion collision experiments have been performed and many observables have been measured with high precision. Later on, several microscopic transport models such as the Boltzmann-Uehling-Uhlenbeck (BUU) model \cite{Bertsch:1988ik} and the quantum molecular dynamics (QMD)\cite{Aichelin:1991xy} and their relativistic versions have been developed. The extraction of the nuclear EOS stood out as one of the primary motivations for HIC studies\cite{Gutbrod,reisdorf1997}. The collective flow and particle (e.g., $\pi$ and kaon) productions are two of the main observables used to extract $K_0$\cite{Molitoris:1986pp,Molitoris:1985gs,Kruse:1985hy,Aichelin:1986ss,Stoecker:1986ci,Cassing:1990dr}. However, many model simulations have demonstrated that the collective flow is sensitive both to the EOS and to the in-medium nucleon-nucleon cross section. It is known that nucleon-nucleon cross sections will be modified by the nuclear medium, however, the details of this modification are still not clear, making the EOS extraction more complicated, see, e.g. Refs\cite{Zheng:1999gt,Persram:2001dg,Andronic:2004cp,Gaitanos:2004ic,Li:2005jy,Zhang:2006vb,BALi08,Li:2011zzp,Kaur:2016eaf}.

In Ref.\cite{Partlan:1994vs}, measurements on the directed collective flow of protons and fragments for Au+Au collisions at beam energies ranging from 0.25\emph{A} to 1.15\emph{A} GeV were presented.
There it was found that neither a soft (represents a value of  $K_0$=200 MeV) nor a stiff (represents a value of  $K_0$=380 MeV) EOS in the QMD model is able to reproduce directed flow data over the entire energy range.
By comparing transport model (pBUU) calculations to the directed and elliptic flows in HICs at the beam energies ranging from 0.15\emph{A} to 10.0\emph{A} GeV, the most extreme $K_0$ (less than 167 MeV or larger than 380 MeV) for EOS were ruled out by Danielewicz et al.\cite{Danie02}. Besides using the collective flow, the kaon yield produced in HICs has also been used to constrain $K_0$. Through a comparison of QMD model simulations to the KaoS data, it was found that simulations with the soft EOS ($K_0$ less than 200 MeV) are superior in reproducing the kaon yields and yield ratios\cite{Hartnack:2005tr,Feng:2011dp}.
Recently, a new observable named $v_{2n}$ regarding the elliptic flow in a broader rapidity range has been presented as a robust probe to constrain the EOS. By comparing the isospin quantum molecular dynamics (IQMD) calculations to the FOPI data, a incompressibility $K_0=190 \pm 30$ MeV was extracted\cite{Fevre:2015fza}.

According to the present study of transport models, a comparison in which the same physical inputs are required, results from the 18 commonly used transport codes are still diversified~\cite{Xu:2016lue}. In Refs\cite{Guo:2013fka,Guo:2014tua}, both the UrQMD and the isospin-dependent Boltzmann-Uehling-Uhlenbeck (IBUU) models have been adopted to study some isospin sensitive observables (such as, pion yield ratio, the yield and flows of nucleons), and it was found that the results visibly depend on the model employed. Thus, a detailed study of the sensitivity of $v_{2n}$ to the EOS and of the effects of various model parameters and their uncertainties seem quite necessary.


\label{sec:1}

\begin{figure}[htbp]
\centering
\includegraphics[angle=0,width=0.40\textwidth]{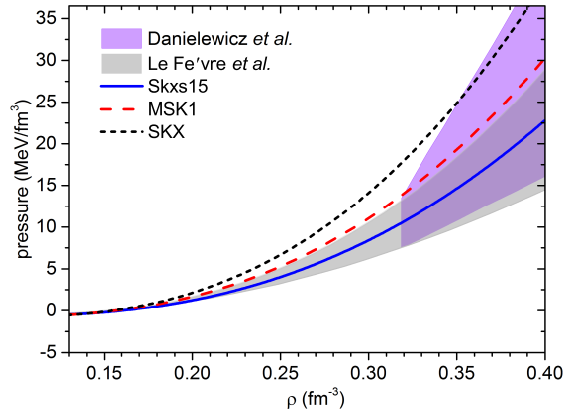}
\caption{\label{fig1}(Color online) Pressure in symmetric nuclear matter as a function of density. The lines represent predictions for the Skxs15 (solid line), MSK1 (dashed line), and SKX (dotted line) interactions. The shaded regions represent the results obtained by Danielewicz et al.\cite{Danie02} and Le F\`evre et al.\cite{Fevre:2015fza}.}
\end{figure}

\begin{table}[htbp]
\centering
\caption{\label{tab:table1} Saturation properties of nuclear matter as obtained with
selected Skyrme parameterizations used in this work.}
\setlength{\tabcolsep}{1.4pt}
\begin{tabular}{lccccccc}
&$K_0$ (MeV)
&&$S_0$ (MeV)
&&$L $ (MeV)\\
\hline
Skxs15 &201 &&31.88&&34.79 \\
MSK1 &234&&30.00&&33.92\\
SKX &271&&31.10&&33.18\\
\hline
SV-sym34 &234&&34.00&&80.95\\

\end{tabular}

\end{table}

To permit a better description of the recent experimental data in HICs at intermediate energies, the Skyrme potential energy density functional has been introduced into the mean-field potential part of the UrQMD code. It is found that with an appropriate choice of the in-medium nucleon-nucleon cross section, the recent published experimental data can be reproduced fairly well\cite{Wang:2013wca,wyj-sym}. In this work, the Skxs15, MSK1, and SKX interactions are chosen which give quite similar values of nuclear symmetry energy (the symmetry energy coefficient $S_0$ and the slope parameter $L$) but the incompressibilities $K_0$ varies from 201 MeV to 271 MeV (e.g., see Table I)\cite{Dutra:2012mb}. It should be pointed out that, the slope parameter $L$ given by the three selected interactions is approximately 34 MeV, i.e., smaller than the average value of 59 MeV\cite{Li:2013ola}. Because the main purpose of this work is to study $K_0$ from elliptic flow, with the introduction of the Skyrme potential energy density functional, $K_0$ and $L$ can not be varied independently, thus these three Skyrme interactions are chosen. In order to examine whether the slope parameter $L$ could affect the results, the SV-sym34 parametrization with incompressibility $K_0=234$ MeV and the $L =80.95$ MeV is also adopted. The results for the pressure in symmetric nuclear matter as a function of density for the Skxs15, MSK1, and SKX interactions are illustrated in Fig.1. For comparison, constraints obtained by Danielewicz et al.\cite{Danie02} and by Le F\`evre et al.\cite{Fevre:2015fza} are also shown with shaded bands. The pressure predicted by SKX lies close to the upper limit of the result of Danielewicz et al.\cite{Danie02}, and the pressure given by Skxs15 lies roughly in the center of the two bands.

Besides the mean field potential part, the in-medium nucleon-nucleon cross section in the collision term, which is still not well-established, also noticeably affects the collective flow. Based on our previously studies~\cite{Wang:2013wca,wyj-sym}, it is found that, by considering a density- and momentum- dependent reduction factor on the free nucleon-nucleon elastic cross section, i.e., the so-called FU3FP4 parametrization, the collective flow and the stopping power data in Au+Au collision at intermediate energies can be reproduced quite well and better.
The FU3FP4 set is, therefore, also here adopted as the preferred parametrization. To show how the in-medium nucleon-nucleon cross section affects the incompressibility, the FU3FP5 parametrization, which represents a stronger reduction of the in-medium elastic cross section, is also considered. Details about the FU3FP4 and FU3FP5 parametrizations of the in-medium elastic nucleon-nucleon cross section can be found in our previous publication\cite{Wang:2013wca}.


\begin{figure}[htbp]
\centering
\includegraphics[angle=0,width=0.48\textwidth]{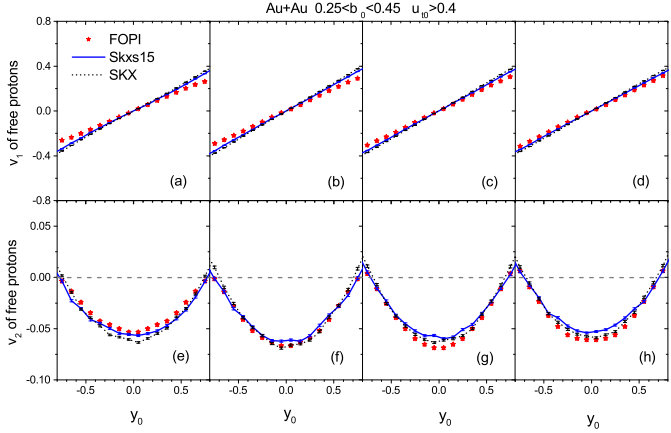}
\caption{\label{fig2}(Color online) The directed flow $v_1$ (upper panels) and elliptic flow $v_2$ (lower panels) of free protons produced in $^{197}$Au+$^{197}$Au collisions at $E_{\rm lab}=0.4A$~GeV (a,e), $0.6A$~GeV (b,f), $0.8A$~GeV (c,g), and $1.0A$~GeV (d,h) with centrality $0.25<b_0<0.45$ and the scaled transverse velocity $u_{t0}>0.4$.
Results calculated with Skxs15 (solid line) and SKX (dashed line) together with the FU3FP4 parametrization of the in-medium nucleon-nucleon cross section are compared with the FOPI experimental data (stars).  }
\end{figure}

Figure \ref{fig2} illustrates the agreement between the UrQMD model calculations and the measured data. Shown are the directed ($v_1=\langle\frac{p_x}{\sqrt {p_x^2+p_y^2}}\rangle$) and the elliptic flow
($v_2=\langle\frac{p_x^2-p_y^2}{p_x^2+p_y^2}\rangle$) of free protons from Au+Au collisions at beam energies of 0.4\emph{A}, 0.6\emph{A}, 0.8\emph{A}, and 1.0\emph{A} GeV, as a function of the normalized rapidity $y_0$. The intervals of the reduced impact parameter $b_0$ and the scaled transverse velocity $u_{t0}$ are chosen to be the same as in the FOPI analysis\cite{FOPI:2011aa}, i.e., $0.25<b_0<0.45$ and $u_{t0}>0.4$, respectively. These quantities are defined as $y_{0}=y_{z}/y_{pro}$ with $y_{pro}$ being the projectile rapidity in the center-of-mass system, $b_0=b/b_{max}$ with $b_{max} = 1.15 (A_{P}^{1/3} + A_{T}^{1/3})$~fm, $u_{t0}\equiv u_t/u_{pro}$ with $u_t=\beta_t\gamma$ the transverse component of the four-velocity and $u_{pro}$ is the velocity of the incident projectile in the center-of-mass system\cite{FOPI:2011aa}. Experimentally, the so-called ERAT method is applied to determine centrality, more detailed discussions of this method can be found in Refs.~\cite{asyeos,Andronic:2006ra,Reisdorf:1996qj}. Theoretically, the impact parameter for each individual event is set at the initialization and can be precisely known. Alternatively, one also can apply the experimental method to determine impact parameter for theoretical simulations. In present work, the ERAT method was applied for simulations of the IQMD model, while the true impact parameter was used for simulations of the UrQMD model. As discussed in Ref.~\cite{Reisdorf:1996qj}, the difference in side flow between simulations with the true impact parameter and the ERAT method is very small. Thus, the determination of centrality will not significantly change our results.

A good agreement between the model calculations and the FOPI data in the whole inspected rapidity range can be seen. Both the slope of $v_1$ and the absolute value of $v_2$ at mid-rapidity calculated with SKX, i.e., with a stronger repulsive potential, are larger than that with Skxs15 but the differences are small. At mid-rapidity, the elliptic flow $v_2$ calculated with SKX is more negative than that calculated with Skxs15; it means protons are more preferentially emitted perpendicular to the reaction plane in the SKX case. Towards the target or projectile rapidity ($y_0$=1 or -1), the $v_2$ calculated with SKX is more positive and crosses the zero line at smaller absolute values of $y_0$ than with Skxs15, indicating that the sideways deflection of the spectator matter for SKX is stronger than that for Skxs15. The reason is that, at the mid-rapidity (the overlapping region of the target and  projectile), the expanding participant matter will be blocked by the spectator matter and will preferentially be squeezed out in directions perpendicular to the reaction plane. A higher pressure generated by the higher incompressibility of SKX leads to a stronger expansion and, consequently, a more negative $v_2$. Around the target or projectile rapidity (spectator nucleons), a higher pressure with SKX causes a stronger deflection of spectator nucleons in the reaction plane resulting in a larger transverse momentum $p_x$ and thus to larger coefficients $v_1$ and a more positive $v_2$\cite{LeFevre:2016vpp}. It has been found that both the directed and elliptic flow at intermediate energies exhibit approximately the scaling behavior, see, e.g., Refs.~\cite{Bonasera:1987zz,Bonasera:1988ewj,Lambrecht:1994cp}. The newly measured high-precision collective flow data provides a new opportunity to investigate the scaling behavior. Further studies are certainly required to understand the physics behind the scaling deviations.


In addition, it has been found that the $v_2$ as a function of rapidity $y_0$ can be well described by a quadratic fit $v_2=v_{20} + v_{22}\cdot y_0^2 $. According to the calculated results shown in Fig.\ref{fig2}, it can be inferred that a stiff EOS leads to larger values of both $|v_{20}|$ and  $|v_{22}|$. Thus the sensitivity to the EOS can be enhanced by using the observable $v_{2n}=|v_{20}| + |v_{22}|$, as discussed in Ref.\cite{Fevre:2015fza}.


\begin{figure}[htbp]
\centering
\includegraphics[angle=0,width=0.4\textwidth]{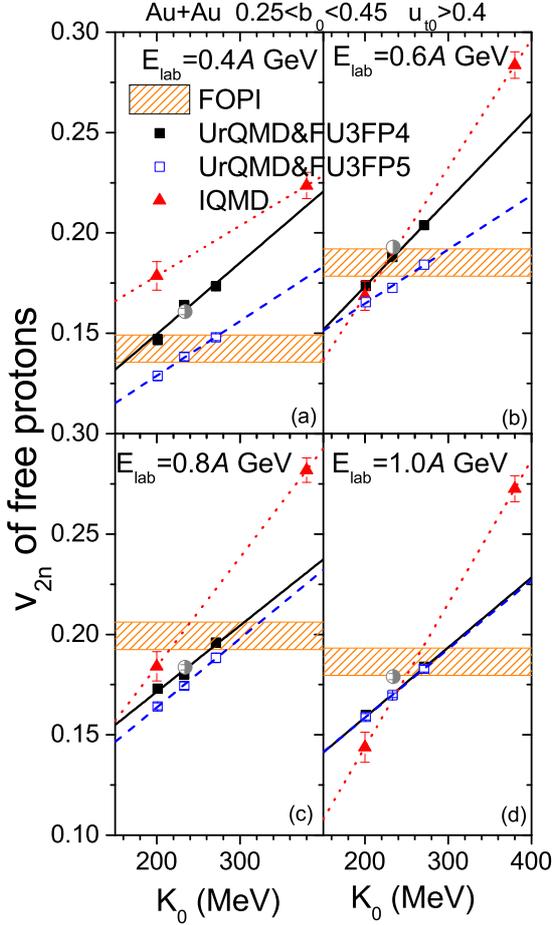}
\caption{\label{fig3}(Color online) The $v_{2n}$ of free protons produced from $^{197}$Au+$^{197}$Au collisions at $E_{\rm lab}=0.4A$, $0.6A$, $0.8A$, and $1.0A$~GeV are shown as a function of the incompressibility $K_0$. In each plot, the shaded bands indicate the FOPI experimental data and full triangles exhibit results from the IQMD model, taken from Ref.\cite{Fevre:2015fza}. Sets of three full squares (open squares)
denote respectively the UrQMD calculations using Skxs15, MSK1, and SKX together with the FU3FP4 (FU3FP5) parametrizations for the in-medium nucleon-nucleon elastic cross section, while the lines represent linear fits to the calculations. Half-solid circles denote calculations with the SV-sym34 representing a force with the incompressibility $K_0=234$ MeV and the slope parameter of the nuclear symmetry energy $L =80.95$ MeV. }
\end{figure}

The $v_{2n}$ of free protons produced in $^{197}$Au+$^{197}$Au collisions at $E_{\rm lab}=0.4A$, $0.6A$, $0.8A$, and $1.0A$~GeV are shown in Fig.\ref{fig3}. Calculations with Skxs15, MSK1, and SKX interactions are compared to the FOPI experimental data, as well as to the IQMD calculations, taken from Ref.\cite{Fevre:2015fza}. First, it can be seen that, the $v_{2n}$ increases strongly with increasing $K_0$ in both the IQMD and UrQMD model calculations, implying that the $v_{2n}$ is indeed sensitive to the incompressibility $K_0$, though this slope dependence is not exactly the same for the two models.. At $0.4A$ GeV, the values of $v_{2n}$ calculated with the IQMD model are significantly larger than that of the UrQMD model, and the difference steadily decreases with increasing beam energy. Reasons for this will be discussed later.

Second, the results of the UrQMD model exhibit an approximate linearity between the $v_{2n}$ and the incompressibility $K_0$. The intersections between the lines and the shaded bands provide the range of expectation for the incompressibility $K_0$. The $v_{2n}$ calculated with the FU3FP4 and FU3FP5 parametrizations are well separated at $0.4A$ GeV, but overlap quite well at $1.0A$ GeV.
The main difference between the FU3FP4 and FU3FP5 parametrizations
consists in a larger reduction of the nucleon-nucleon elastic cross section at lower
 relative momenta in FU3FP5 than in FU3FP4. The reduction factor for both FU3FP4 and FU3FP5 approaches unity at higher relative momentum~\cite{Wang:2013wca}, hence at higher incident energies.

On average, the central value of the incompressibility $K_0$ is obtained to be 240 MeV for calculations with the FU3FP4 parametrization, while it reaches 275 MeV for FU3FP5. Those results are larger than that from the IQMD model simulations using the same observable, which is about 222 MeV (evident also from Table 2 in Ref.\cite{Fevre:2015fza}). The main difference comes from the collision term in the two models, i.e., the free nucleon-nucleon cross section is used in the IQMD model, while a density- and momentum-dependent in-medium nucleon-nucleon cross section is used in the UrQMD model. The difference between the two model calculations become smaller at higher beam energies. This can be understood from the near equivalence of the in-medium and free cross sections at the higher relative momenta prevailing at higher beam energies. The remaining difference between the two models may stem from (I) different treatments in the Pauli blocking which also determine the collision rate. It can be seen from the transport model comparison paper I\cite{Xu:2016lue} that the Pauli blocking rate in the IQMD model is higher than that in the UrQMD model. Therefore, a larger sensitivity of $v_{2n}$ to the incompressibility $K_0$ is seen in the IQMD model. They may further stem from(II) different values of the width of the Gaussian wave packet, as well as different parameters used in the cluster recognition criteria. Influences of those treatments on the $v_{2n}$ deserve further studies.  With a weaker reduction of the in-medium nucleon-nucleon cross section, the extracted $K_0$ will be smaller. It may explain the reason why the $K_0$ obtained from the IQMD model is smaller than that from the UrQMD model. $K_0 = 240 \pm 20$~MeV ($K_0 = 275 \pm 25$~MeV) for the FU3FP4 (FU3FP5) parametrization of the in-medium nucleon-nucleon cross section, which best describes the experimental data, can be extracted within a 2-$\sigma$ confidence limit from the chi-square test.

Third, we note that with both models, the value of extracted $K_0$ increases with increasing beam energy, e.g., $K_0=180\pm20$ MeV is favored at the beam energy of $0.4A$ GeV while $K_0=280\pm17$ MeV fits best to the data taken at $1.0A$ GeV in the UrQMD model description with FU3FP4. It is known that the emitted free protons are sensitive on both the maximum densities reached in the collision and to the contributions of inelastic channels (mainly the Delta degree of freedom), which become larger with an increasing incident energy. If these dependencies are not fully reproduced by the transport model and the employed Skyrme forces, e.g., the high orders of the EOS dependence on density like the skewness, they may result in an energy dependence of the deduced $K_0$ parameter as it is observed here.

Further, one sees clearly that the results for $v_{2n}$ calculated with SV-sym34 and MSK1, i.e., with the same value of $K_0$, are very close to each other even though the difference in $L$ is as large as 47 MeV. It illustrates the $v_{2n}$ is much more sensitive on the nuclear incompressibility than on the nuclear symmetry energy. Because the nuclear symmetry potential is relatively weak compared to the isoscalar part of the nuclear potential, its weak effect on observables is not easy to see, usually, the difference or ratio between isospin partners can provide some hints for the isovector part of the nuclear potential.

\begin{figure}[htbp]
\centering
\includegraphics[angle=0,width=0.4\textwidth]{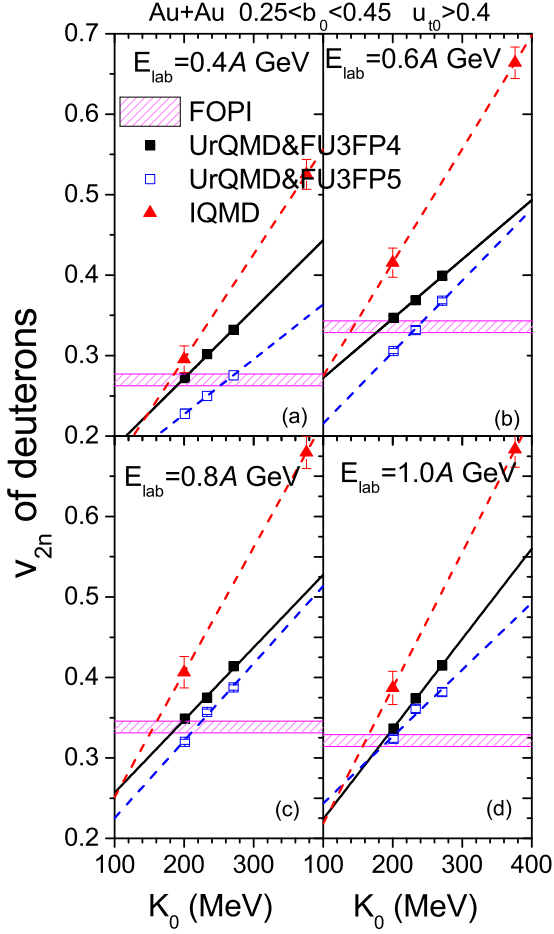}
\caption{\label{fig4}(Color online) The same as Fig.\ref{fig3} but for the $v_{2n}$ of deuterons.  }
\end{figure}

Similarly to Ref.\cite{Fevre:2015fza}, we were interested in constraining further $K_0$, looking at flows of emitted light isotopes. Here we focus on the elliptic flow of deuterons, which rapidity dependence is well reproduced by the UrQMD approach, still with the FU3FP4 parametrization. $v_{2n}$ is equally well described, as shown in Fig.\ref{fig4} for the same Au+Au collisions at beam energies between 0.4\emph{A} and 1.0\emph{A} GeV. The correlation of this quantity with the incompressibility $K_0$ is again linear in the description with the UrQMD model. As with the protons, the differences between the compressibilities obtained with the two choices for the in-medium cross section diminish with increasing bombarding energy. Unlike with the free protons, the expectation range of $K_0$ here is weak dependent on the incident energy, like found in Ref.\cite{Fevre:2015fza} with the IQMD predictions. By averaging over the four energies, we derive $K_0 = 190 \pm 10$~MeV ($K_0 = 225 \pm 20$~MeV) for the FU3FP4 (FU3FP5) parametrization, results that are larger than the $K_0 = 170 \pm 8$~MeV deduced with the IQMD model. In both models, the value of $K_0$ extracted from the $v_{2n}$ of free protons is about 50 MeV larger than that of deuterons. This difference is approximately of the same order as the energy dependence of $K_0$ observed for protons that, in contrast, is not observed for deuterons. Calculations performed with the TuQMD transport model have indicated in Ref.\cite{asyeos} that quite higher densities are probed by free protons than by composite light particles. In this context, over our present beam energy systematics, deuterons probe a lower and narrower range of densities than the protons. Therefore they are less sensitive on the inaccuracy of the employed Skyrme potentials in describing the high order terms of the density dependent EOS for a given parametrization, i.e., a given apparent $K_0$. Overall, however, the analysis of the rapidity dependence of the elliptic flow supports the soft choice for the symmetric-matter equation of state, up to two times the saturation density as deduced for studies of Refs\cite{Fevre:2015fza,asyeos}.  It is known that the higher order term (such as the skewness coefficient $Q_0$) may be required for a correct description of the EOS at high densities (such as 2$\rho_0$)\cite{Chen:2009wv}. In one-parameter descriptions as attempted here, the parameter $K_0$ thus represents an effective incompressibility that includes effects of the higher terms in the density range that is tested. On one hand, one should not expect that the $K_0$ (as used to characterize the EOS in transport models) obtained from HICs is directly comparable to the one deduced from nuclear properties (e.g., GMR), since there is no basic principle which requires that the parabolic approximation of EOS is valid over large ranges of density. However, on the other hand, in the case of the Skyrme and Gogny energy density functionals, the incompressibility $K_0$ and the skewness coefficient $Q_0$ are well correlated (see, e.g.,\cite{Khan:2013mga,Chen:2011ib,Chen:2009wv}). It is interesting to find that our present result is also consistent with many constraints on $K_0$ extracted from nuclear properties.

  We note here that, by varying the in-medium nucleon-nucleon cross section, some other observables, for instance the yield of hard photons, also vary to some extent\cite{Cassing:1990dr,Nifenecker:1990um,Migneco:1993nhc,Bonasera:2006qb,Yong:2011nz,Ma:2012zzb}. It would be of great interest to investigate simultaneously the photon-related observables and nucleon-related observables in the same reaction. Hopefully, in-depth understanding of the in-medium nucleon-nucleon cross section can be achieved by studying different observables.

In summary, by comparing the UrQMD model calculations with the recent FOPI data for the elliptic flow in Au+Au collisions in the beam energy range 0.4\emph{A} - 1.0\emph{A} GeV, it is found that the nuclear incompressibility $K_0$ is quite sensitive to the $v_{2n}$, a quantity obtained from a quadratic fit ($v_2=v_{20} + v_{22}\cdot y_0^2 $) of the elliptic flow as a function of rapidity by adding the coefficients as $v_{2n}=|v_{20}| + |v_{22}| $, and the $v_{2n}$ increases almost linearly with increasing the incompressibility $K_0$. The influences of the in-medium nucleon-nucleon cross section and the nuclear symmetry energy on the $v_{2n}$ are also analyzed. It is found that the $v_{2n}$ can be affected by the in-medium nucleon-nucleon cross section but hardly influenced by the nuclear symmetry energy. With the FU3FP4 parametrization (i.e., the preferred choice in the present version of the UrQMD model) of the in-medium nucleon-nucleon cross section, $K_0 = 240 \pm 20$~MeV and $K_0 = 190 \pm 10$~MeV are extracted from the $v_{2n}$ of free protons and deuterons, respectively. By combining the error intervals of the proton and deuteron results, an averaged $K_0 = 220 \pm 40$~MeV is obtained. The extracted $K_0$ will be smaller (larger) if a weaker (stronger) reduction on the in-medium nucleon-nucleon cross section is used. Additional calculations with other model assumptions and/or transport models will be certainly required to confirm the sensitivity of the $v_{2n}$ to the EOS and assess the obtained result.

\section*{Acknowledgements}
Fruitful discussions with Ch. Hartnack are greatly appreciated. The authors acknowledge
support by the computing server C3S2 in Huzhou University. The work is supported in part by the National Natural
Science Foundation of China (Nos. 11505057, 11375062, 11605270, 11647306, and 11747312), and the Zhejiang Provincial Natural Science Foundation of China (No. LY18A050002).

\end{document}